# Ransomware Analysis using Feature Engineering and Deep Neural Networks

Arslan Ashraf[a], Abdul Aziz[a], Umme Zahoora[a], Muttukrishnan Rajarajan[c] and Asifullah Khan[*a, b]

[a]Department of Computer Science, Pakistan Institute of Engineering and Applied Sciences, Nilore-45650, Islamabad

[b]Centre for Mathematical Sciences, Pakistan Institute of Engineering and Applied Sciences, Nilore-45650, Islamabad

[c] School of Mathematics & Engineering, City University London, London EC1V 0HB, UK

***Abstract*—**Detection and analysis of a potential malware specifically, used for ransom is a challenging task. Recently, intruders are utilizing advanced cryptographic techniques to get hold of digital assets and then demand a ransom. It is believed that generally, the files comprise of some attributes, states, and patterns that can be recognized by a machine learning technique. This work thus focuses on the detection of Ransomware by performing feature engineering, which helps in analyzing vital attributes and behaviors of the malware. The main contribution of this work is the identification of important and distinct characteristics of Ransomware that can help in detecting them. Finally, based on the selected features, both conventional machine learning techniques and Transfer Learning based Deep Convolutional Neural Networks have been used to detect Ransomware. In order to perform feature engineering and analysis, two separate datasets (static and dynamic) were generated. The static dataset has 3646 samples (1700 Ransomware and 1946 Goodware).

On the other hand, the dynamic dataset comprised of 3444 samples (1455 Ransomware and 1989 Goodware). Through various experiments, it is observed that the Registry changes, API calls, and DLL's are the most important features for Ransomware detection. Additionally, important sequences are found with the help of the N-Gram technique. It is also observed that in the case of Registry Delete operation, if a malicious file tries to delete registries, it follows a specific and repeated sequence. However, for the benign file, it doesn't follow any specific sequence or repetition. Similarly, an interesting observation made through this study is that there is no common Registry deleted sequence between malicious and benign files. And thus this discernable fact can be readily exploited for Ransomware detection.

---

## 1. Introduction

Ransomware is a type of malware program that infects, locks or takes control of a system with the intention of extorting money from its owner. Ransomware and its rapidly emerging variants are the hazardous profitable threat. According to a survey conducted in 2016, out of 290 selected organizations, 50% were its victims, and around 40% of its target victims have paid millions of ransom in a year(Morato et al. 2018). Ransomware attack threatens its victims to publish, delete or change their data for ransom, therefore it had become the most dangerous threat for enterprises, Small-Medium Businesses (SMB), and individuals. The Ransomware attack has been around for approximately 2 decades, however, its use to demand ransom has increased recently and was first occurred in the mid-1990s (Sultan et al. 2018). Recently in 2017, Wannacry the most famous and destructive ransomware variant has locked the data and demanded a ransom of about £92 million from many organizations including Britain's National Health Service, some of Spain's largest companies like Telefónica, as well as computers across: Russia, Ukraine, and Taiwan (Ashok Koujalagi 2018). Ransomware attackers are working fearlessly because of the use of bitcoin cryptocurrency and are still unknown. When this type of malicious software unknowingly invades any single host, using known vulnerabilities it can easily infect the whole network.

These wide range damages motivated researchers to explore various technologies to prevent such attacks. (Cabaj et al. 2018) presented SDN (Software Defined Network) based detection approach that exploits HTTP (Hyper Text Transfer Protocol) message sequences and content size to detect Crypto-Ransomware. (Chen and Bridges 2017), uses signature generated by YARA rules augmented with dynamic analysis to detect Wannacry Ransomware attack. To detect different families of ransomware attack existing remedies are based on different discriminative features extracted statically, dynamically or both. Static analysis of malware is performed without executing it. (Schultz et al. 2001) are known as one of the initial people to detect malware by selecting different static features with the help of data mining. He used Portable Executable (PE) header feature, byte sequences, and extracted strings. This method got the detection rate of 97.7%. (Shabtai et al. 2009) worked on the PE file header and extracted static features for the classification of malicious files. He grouped all the PE features in six categories named such as PE header, optional header, import, export, resource, and version number. (Liao) achieved accuracy with 0.2% false positive rate by selecting five fields of PE header and implemented hand-crafted algorithm. (Ijaz et al. 2019) analyzed malware using machine learning techniques like Logistic Regression, Decision Tree, Bagging Classifier, Gradient B. Classifier and achieved 99% AUC-ROC over static malware analysis. Static methods provide quick and safe detection without executing the malware. However, they are unable to detect the polymorphic variations that change their appearance to deceive static methods.

Dynamic Analysis that is based on observing run time behavior of malware in a safe environment is more robust to polymorphic variations. (Young 2006) employed cryptography to carry out extortion-based attacks, by implementing the payload through Microsoft Cryptographic API. Guarnieri et al., proposed a new method of dynamic analysis in a sandbox environment that served as a basis for future research. For Example, (Sgandurra et al. 2016) presented a machine learning: EldeRan approach, which dynamically analyzes and classifies ransomware attack. Chen et al., extracted discriminative features by applying the Term Frequency (TF) and Inverse Document Frequency (IDF) technique on logs generated by Cuckoo Sandbox. (Kumar et al. 2019) used different machine learning approaches like KNN (K-Nearest Neighbor), Logistics Regression, Decision Tree and Random Forest to classify the benign and malicious set. To intercept dynamic analysis-based intrusion detection system attacker introduces metamorphic variations that along with appearance can change functionality. Therefore, it is essential to find more robust features that can cope with these concealment techniques more efficiently.

The objective of this research work is to perform feature analysis in order to find the most discriminative features and sequences that can better classify the benign and attacked samples. For this purpose, RanSD (Ransomware Static and Dynamic Analysis) system is proposed, that follows the procedure as follow:

1. Firstly, generated static and dynamic datasets, by collecting different Ransomware samples from different sources. To generate dynamic data used Cuckoo Sandbox to analyze activities performed by any files in a controlled environment. Similarly, for the static dataset, PE File information was extracted from different executable files with the help of the built-in package of python library named as PE File.
2. By exploiting Wrapper based Mutual Information (MI), the most contributed behaviors in the detection of ransomware were reduced from 40,000 to 300.
3. Then, sequences followed by malicious and benign files were analyzed and some of them were extracted for further detection of maliciousness more efficiently.
4. Finally, passed the most contributing features to the machine learning based detection system. Transfer learning, using a pre-trained model on ImageNet of ReeNeT-18, has been applied along with SVM and Random Forest algorithm.

The rest of the paper is organized as follows: Section 2 discusses the proposed RanSD methodology and procedural details of the dataset generation. Experiments performed and their results on static and dynamic datasets for analyzing the most important behaviors are in section 6. Finally, conclusion and future work related are discussed in section 7.

---

## 2. Proposed RanSD technique

This section provides the details of the material and methods used to collect and analyze samples. Moreover, includes the details of the methodology of feature engineering techniques that assisted in finding the most discriminative features and sequences. Related theory of different detective systems that were trained on extracted features is also part of the discussion. Proposed technique RanSD is based on three modules as described in Figure 1: where the first module is a feature extraction module that is based on sample collection and feature extraction, the second module is a features analysis module that extracts only relevant features and sequences for detection module. In the detection module, the effectiveness of selected features is validated on different conventional learning algorithms and also applied deep learning-based transfer learning.

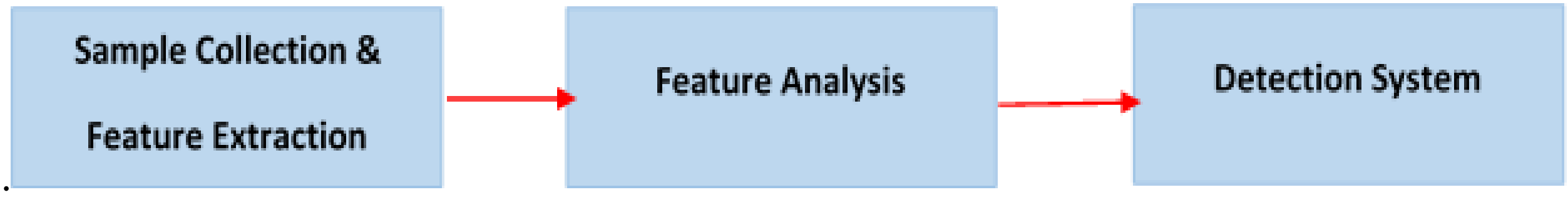

**Figure 1: Abstract diagram of the proposed methodology**

### 2.1 Samples collection and preparation

The very first step for machine learning models and experiments is the availability of datasets. Since the interest of intruders is increasing towards ransomware day by day and they are applying recent and innovative techniques to collect ransom amount and developing more sophisticated approaches to lessen the chances of recovery of data. On the other hand, to ensure data security easier and possible for all users and organizations, it is necessary to develop such a system that can detect and analyze different features and behaviors of malicious files. In order to achieve these goals, about 45,000 malicious samples were collected from different well-known sources e.g. VirusTotal, VirusShare, etc. On the contrary, about 3000 goodware samples were collected from Windows 7 Operating System by handcrafted parser algorithm. This work is composed of both static and dynamic analysis, therefore these collected samples were used for both purposes.

### 2.2 Feature extraction

In this work, the classification of ransomware and benign files is carried out on the basis of both static and dynamic features. PE file information and Cuckoo Sandbox are used for extracting the static and dynamic features respectively.

#### 2.2.1 Static feature extraction

In order to generate the static dataset, 3650 executable files were collected and 89 most contributing features were extracted from PE file structure using a self-designed parser. The parser uses a built-in "pefile" python library to extract required states from different headers which are mentioned in Table 1 and Figure 2. PE File structure is a standard format of Microsoft, whenever any state deviates from its standard value, it performs malicious activity. Some of the main features are SizeOfInitializedData, DLLCharacteristics, SizeOfCode, FileAllignment, SizeOfHeaders, MajorImageVersion, Entropy and Checksum.

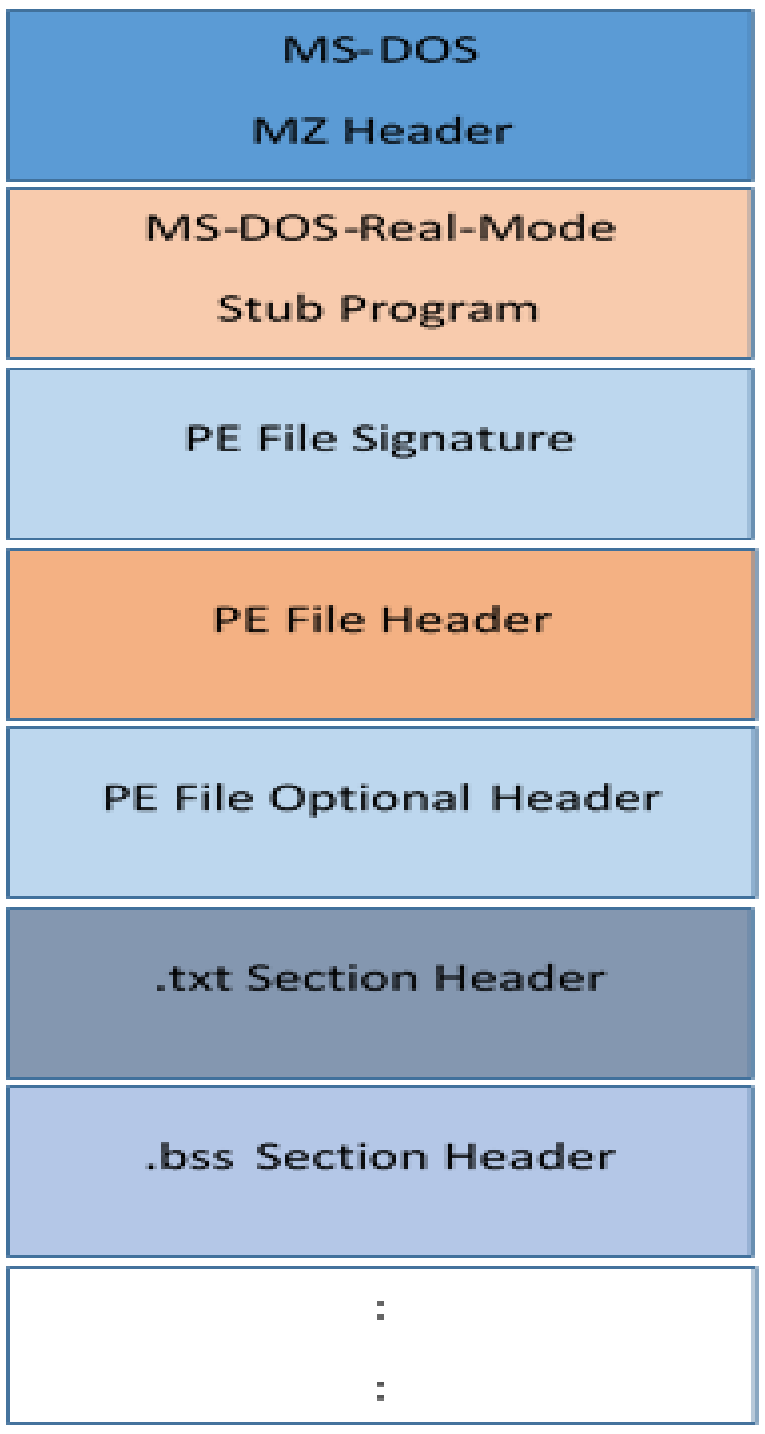

**Figure 2: PE file format**

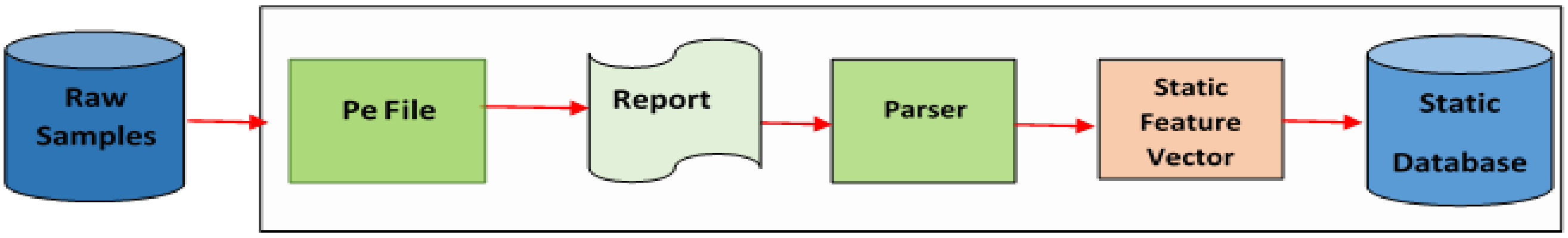

**Figure 3: RanSD static feature extraction process**

**Table 1: Feature vector of static dataset**

| Name of Headers | No. of Features extracted | Percentage |
|---|---|---|
| Optional Header | 26 | 29% |
| Section Header | 12 | 13% |
| DOS Header | 18 | 20% |
| PE File Header | 9 | 10% |
| Directory Entry Resource | 4 | 4% |
| Directory Entry Load Config | 13 | 14% |
| Derived | 9 | 10% |

### 2.2.2 Dynamic feature extraction

Sandboxing is considered as one of the best approaches for dynamic feature extraction, which gives the flexibility to execute and record the behavior of files in a controlled environment. Similarly to generate the dynamic dataset, an emulator was designed which is a controlled environment (Cuckoo Sandbox). In this environment, there can be multiple guests and a host. In our case, this environment is composed of only one guest i.e. Windows7 and Ubuntu as host. Then allowed every file to execute in this controlled environment and recorded all their behaviors in the *Report.json* file. This log file records all the behavior and changes that occur during the execution of a file in the guest machine. Through these records, API calls, Registry operations (Delete, Create, Read, and Write), File operations (Delete, Create, Read, and Write), Directory operations (Delete, Create, Read, and Write), Network Domains, Drop File extensions, Strings and Dll's all were recorded and used as features. About 47,000 of total features were extracted initially, Table 2 is showing the summary details of these features.

**Table 2: Dynamic feature vector**

| Name of Features | Presence of feature | Percentage |
|---|---|---|
| API calls | 241 | 0.51% |
| Registry operations | 8476 | 18% |
| File operations | 31857 | 67.6% |
| Directory created | 1121 | 2.38% |
| Extensions | 1776 | 3.77% |
| Drop | 800 | 1.7% |
| DLL's | 316 | 0.67% |
| Strings | 2496 | 5.3% |

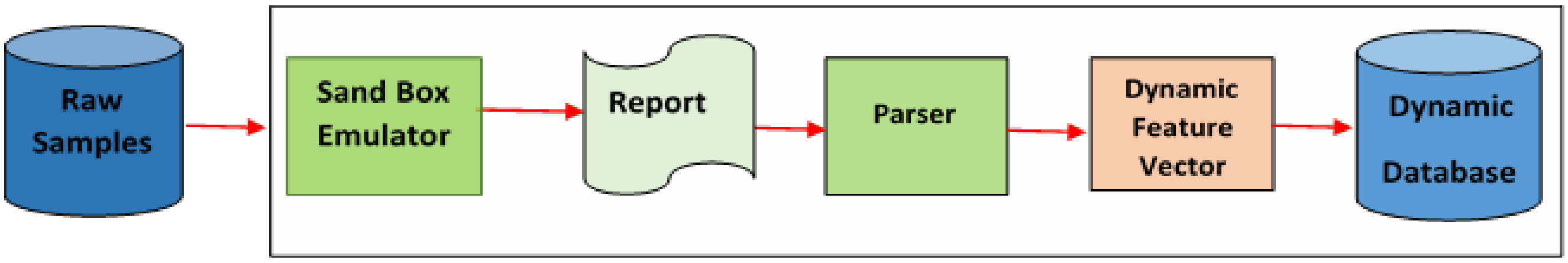

**Figure 4: Dynamic feature extraction**

### 2.3 Feature engineering techniques

The proposed RanSD technique used different features engineering techniques (shown in Figure 5) to reduce the detection time and increase the accuracy. Detailed overview of used Feature Engineering methodologies is provided below.

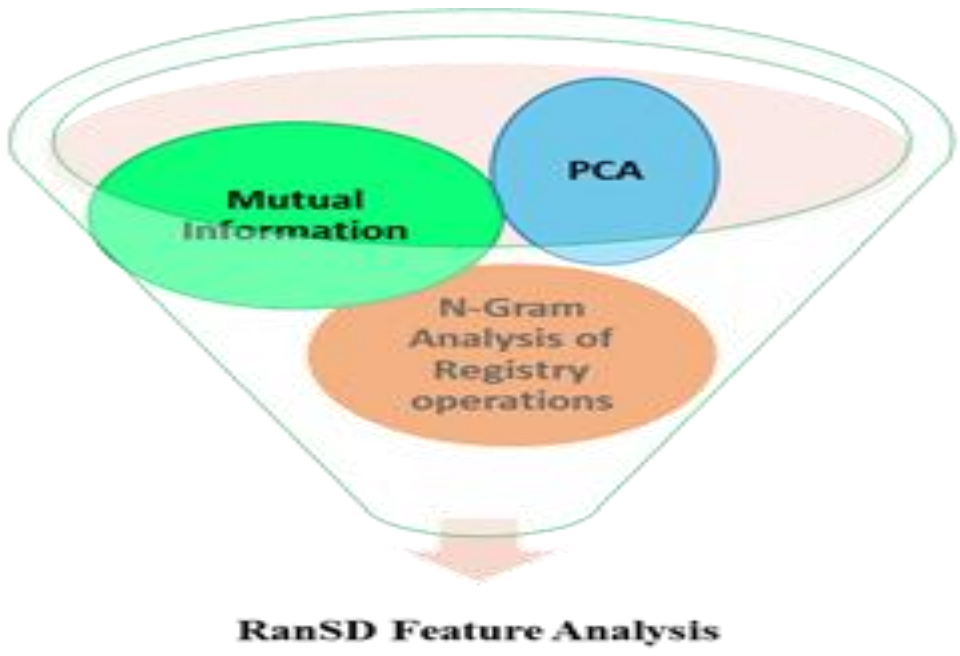

**Figure 5: Feature engineering**

### 2.3.1 Principle component analysis (PCA)

PCA (Abdi and Williams 2010)is one of the most important technique to reduce the dimensionality over a large number of correlated features. It tries to find out the directions in which the maximum variance of the given data is being captured. Eq (1-3) are mathematically describing how to calculate the variance of the projected data along *w*,

$$Var(z) = Var(w^T x) = \frac{1}{N}\sum_{i=1}^{N}\left[(W^T x_i - W^T \mu_x)^2\right] \tag{1}$$

$$= w^T \frac{1}{N}\sum_{i=1}^{N}\left[(x_i - \mu_x)(x_i - \mu_x)^T\right]^w \tag{2}$$

$$= w^T C w \tag{3}$$

Taking derivative with respect to $w^T$ and substituting '0' yields, $Cw = \alpha w$. This is an Eigen vector problem. $C$ is covariance matrix, ***'w'*** corresponds to Eigen vector, $\alpha$ is the Eigen value of that Eigen vector. So, these Eigen vectors are principle components which are in the direction of maximum variance.

### 2.3.2 Mutual information (MI)

MI (Peng et al. 2005) is the dimensionless quantity that measures how much one variable gives us the information about another variable. The joint probability distribution of two discrete variables *'x 'and 'y '* is $P_{XY}(x,y)$ and MI in between them is $I(x;y)$ and is mathematically describe in Eq. 4. Here $P_X(x)$ and $P_Y(y)$ are the marginal distributions. Where *'E'* (in Eq. 5) defines the expected distribution over 'P'.

$$I(X;Y) = \int_X \int_Y P_{XY}(\mathrm{x},\mathrm{y}) \log \frac{P_{XY}(x,y)}{P_X(x)P_Y(y)} \tag{4}$$

$$= E_{P_{XY}} \log \frac{P_{XY}}{P_X P_Y} \tag{5}$$

MI finds the similarity between the joint distribution and the product of factored marginal distribution. If *'x'* and *'y'* are independent, then the joint distribution of *'x'* and *'y'* will be equal to the product of factored marginal distribution. Further solving integral over the product of factored marginal distribution will be evaluated to zero which shows that the two variables are unrelated. In feature selection task objective is to maximize the MI between the set of *'k'* selected attributes and the target label *'y'*. Where *'k'* is showing the number of features selected, finding the value of *'k'* is an NP-hard problem, Where RanSD MI technique solved this issue by using two methods that validate each other. Firstly use the PCA scree plot to determine the number of principal components that can capture variance and then validated it further by performing a greedy stepwise wrapper based feature selection algorithm that incrementally adds the feature to find its importance in detection, for detection purpose SVM is evaluated using accuracy matrix.

### 2.3.3 N-gram technique

N-gram model is the probabilistic model for natural language processing which helps in finding the sequences over an order of *(n – 1)*. N-gram (Kešelj et al. 2003)distribution is used to model speech recognition, a combination of words or characters in any type of data. Where 'N' represents the number of features taken in to account for finding any sequence. If $N$=2 then it is known as Bigrams and it is useful for finding all pairs of words in any document collection. For example, a sentence "Please take the notice" will be paired as "Please take", "take the", "the notice". Similarly for a 3-gram analysis, all possible sequences of three words will be considered.

RanSD Feature engineering module employed N-gram on Registry Operations and compared it against benign sequences. The objective of employing N-gram was to find out the underlying discriminant pattern that tends the MI feature engineering technique to select Registry Operations as the most contributing attributes. Important underlying sequences are extracted by finding the probability against all Registry Operations in a way, if one Registry operation has occurred, what is the probability other operations will also occur? It brings forth very interesting results that have been discussed in detail in the results and discussion section.

### 2.4 RanSD detection system

In order to determine the detection performance of selected features on different decision boundaries, multiple classifiers have been trained and tested including SVM, Random Forest, and ResNet-18. ResNet-18 neural network architecture has been applied for transfer learning. This architecture has already been trained on ImageNet. Results show that selected features are not only good for SVM that is used as an evaluator in wrapper feature selection but can also well generalize the other diverse decision boundaries. Tested decision surfaces include an ensemble of the decision trees as a base classifier, hyperplane based decision surface, and representation learning based transfer learning. Brief description of the classification methodologies of these models is given below.

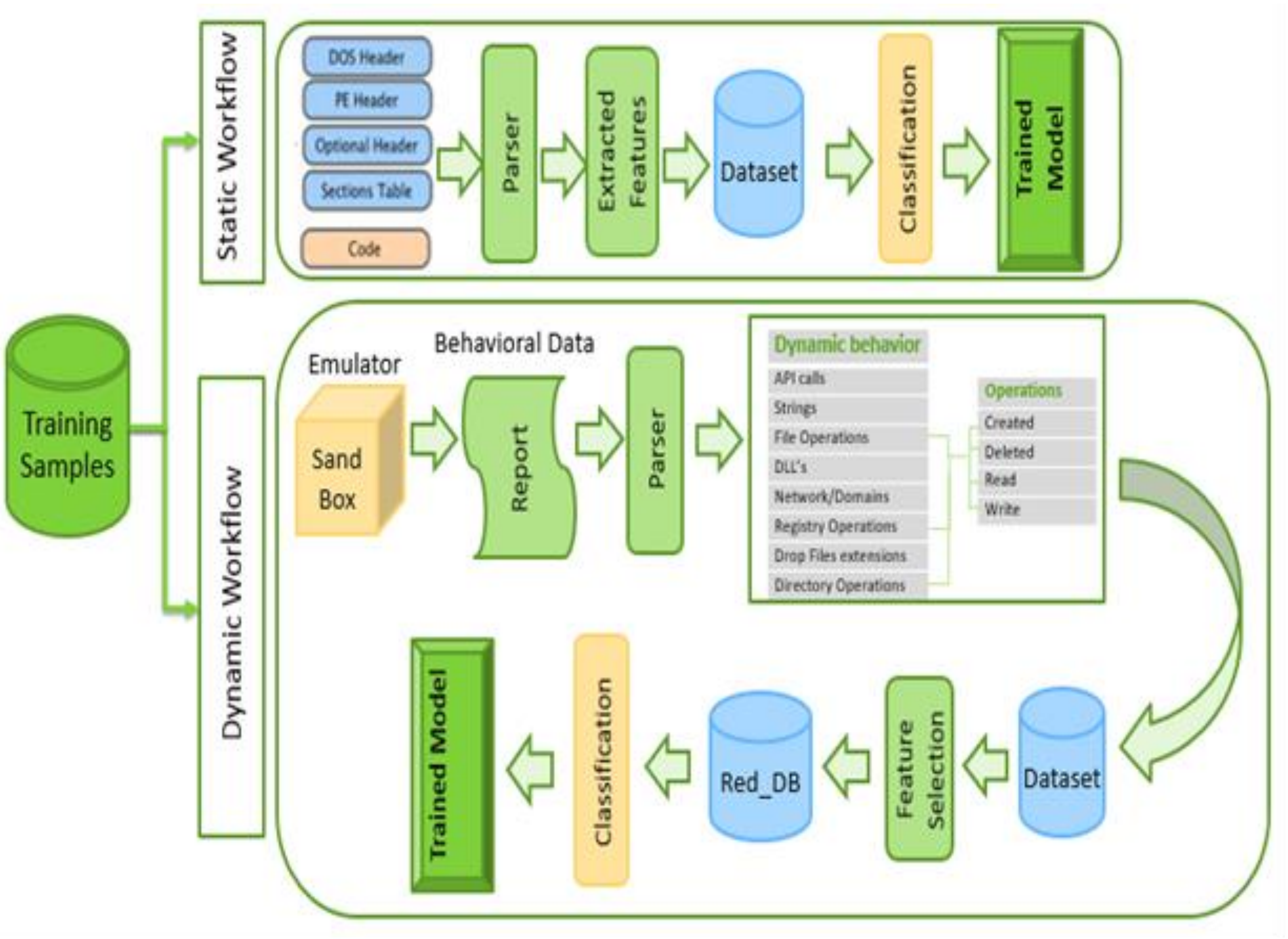

**Figure 6: Overview of proposed methodology**

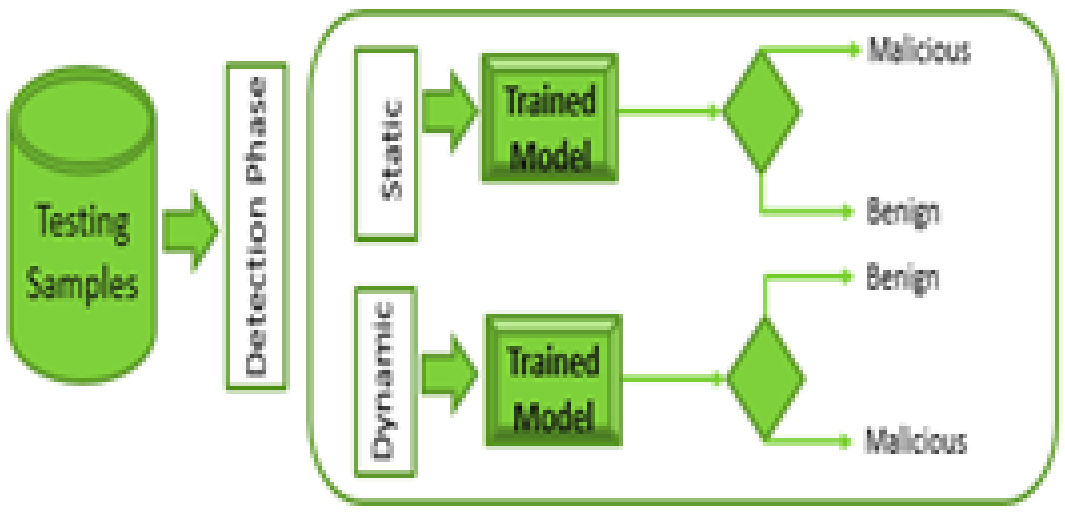

**Figure 7: Testing Phase**

### 2.4.1 Conventional classification system

#### 2.4.1.1 Random forest (RF)

RF technique (Breiman 2001)is an ensemble based method that combines the results of multiple decision tree classifiers for a single output. RF contributed an extra functionality in the begging process, by drawing randomly bootstrap samples from training data. Further, each selected sample grows as an unpruned tree and split decision at each node is based on best selected predictor from randomly chosen predictors. In the testing phase, the label is assigned on the basis of a majority decision by base classifiers.

#### 2.4.1.2 Support vector machine (SVM)

SVM is a linear separator for high dimensional data. It (Scholkopf and Smola 2001) linearly separates two classes by locating hyperplane with maximum margin from nearby few points of opposite classes. These points are also known as support vectors. Soft margin SVM classifier allows misclassification with a penalty that is often denoted by *'C'*. When *'C'* is small, it imposes a little penalty and allows some misclassification. While when *'C'* is large, it imposes a greater penalty and optimizer tries to avoid misclassification.

### 2.4.2 Deep learning based classification system

#### 2.4.2.1 ResNeT deep neural network

Residual Networks (ResNet) (He et al. 2016) is a model neural network used for many computer vision tasks. This model was the champion of ImageNet challenge in 2015. The major contribution of the ResNet was to train extremely deep neural networks with 150+layers successfully. Earlier to ResNet training very deep neural networks were challenging owing to the problem of vanishing gradients. Residual learning incapacitates the vanishing gradient problem using skip connection. Skip connections work much better than other multiplicative interactions. As builds a deeper network by simply repeating the module. Residual learning eases optimization, and decreases the error rate for deeper networks.

#### 2.4.2.2 Transfer learning

In literature, many researchers are exploiting the deep representations for transfer learning across the domains. Uses of transfer learning (Khan et al. 2019b; Qureshi et al. 2019) (Rezende et al. 2017; Kim et al. 2018) and deep learning (Hardy et al. 2016; Kolosnjaji et al. 2016; Saxe and Berlin 2016; Chouhan et al. 2019) for malware detection in literature, motivated us to practice it as a ransomware detection system. Transfer learning facilitates to use of the knowledge of an already trained machine learning model to a different but allied problem. It exploits, what has been learned in one domain, to develop generalization in an alternative domain. Knowledge is being handover in terms of weights that a network has learned at domain A to a novel domain B. Some pre-trained machine learning models out there became quite popular. One of them is the ResNet model used in the proposed RanSD detection module, which was trained for the ImageNet "Large Visual Recognition Challenge".

## 3. Implementation details

Experimental setting for dataset generation is already been discussed in Section 2. Whereas remaining experiments and analysis has been performed on Python 3.5, Ubuntu Operating System installed on Intel Core i7-6700 CPU@ 3.40GHzx8, GeForce GTX 1070/PCIe/SSE2 system having memory 61.1GB.

## 4. Parameter settings

Overall datasets were split into two divisions, 80% for training and validation and rest 20% left for the test. To demonstrate the performance over the unseen dataset, we performed cross-validation with 5-folds to select the models. The Parameter setting for ResNet-18 is given in Table 3.

**Table 3:Parameters Setting**

| Parameter Settings | |
|---|---|
| Architecture | ResNet-18 |
| Previously Trained | ImageNet Dataset |
| Criterion Loss | Cross Entropy Loss |
| Optimizer | Stochastic Gradient Descent |

## 5. Performance evaluation measures

Proposed RanSD technique generated balance dataset for binary class problem, henceforth uses accuracy as performance measure to evaluate RanSD detection models. Additionally, evaluated models on the basis of precision. Mathematical description of accuracy, precision, and recall are described below.

$$Accuracy = {TP + TN}/{TP + FP + FN + TN} \tag{6}$$

$$\text{Recall} = TP/TP + FN \tag{7}$$

$$\text{Pr}ecision = TP/TP + FP \tag{8}$$

## 6. Results and discussion

Proposed RanSD technique is based on static and dynamic feature analysis using PCA that was discussed in detail in Section 6.1.1. Further exploited Wrapper based Feature Selection technique to extract the most contributing dynamic behavior for ransomware detection. Where, Registry Operations were selected as the top contributing operations. Experiments include the n-gram sequence analysis of Registry Operation to find the underlying discriminative patterns. Furthermore, to validate the selected features different learning models were trained and their test performance is shown in Tables 4, 5 and 6.

## 6.1 Feature analysis & reduction

### 6.1.1 Principle component analysis (PCA) of static and dynamic data

PCA was applied to check the covariance of features over the static data and to reduce the dimensions with maximum variance. With the help of PCA, static dataset reduced from 89 dimensions to approx. 56 dimensions as shown in Figure 8.1. Similarly, on applying PCA over 47,000 features of dynamic dataset, it is observed that about 99.9 % of variance is being captured in about 300 features as shown in Figure 8.2

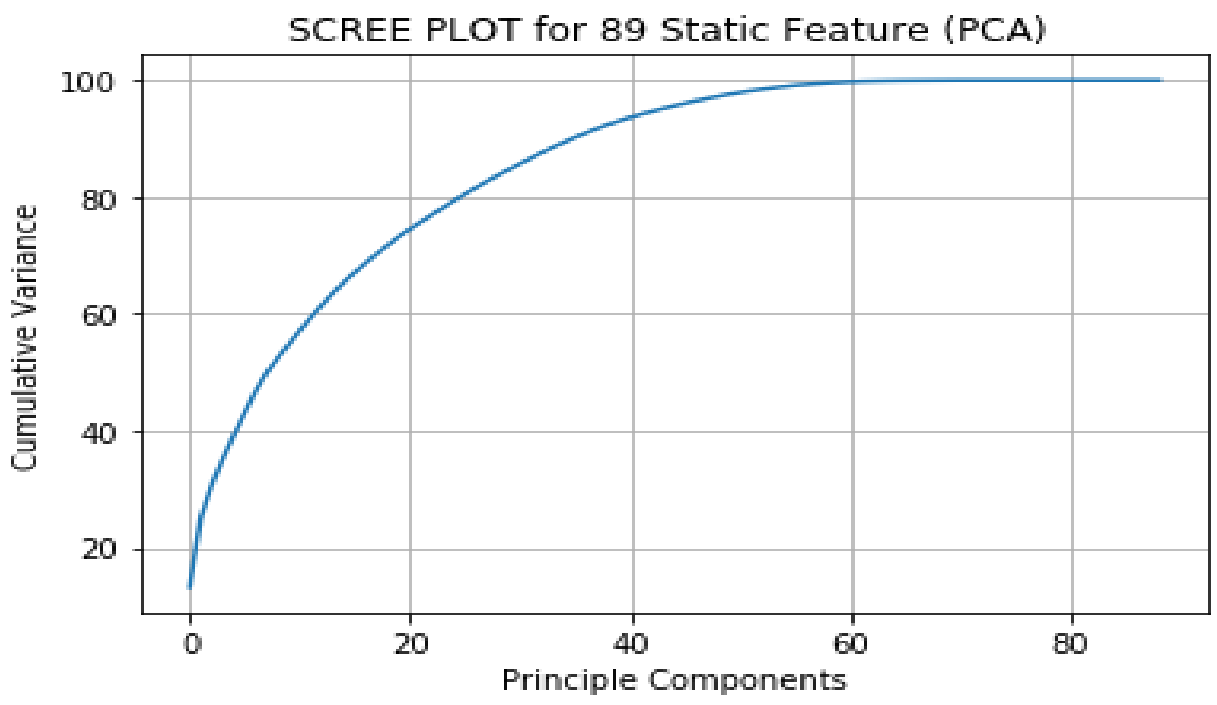

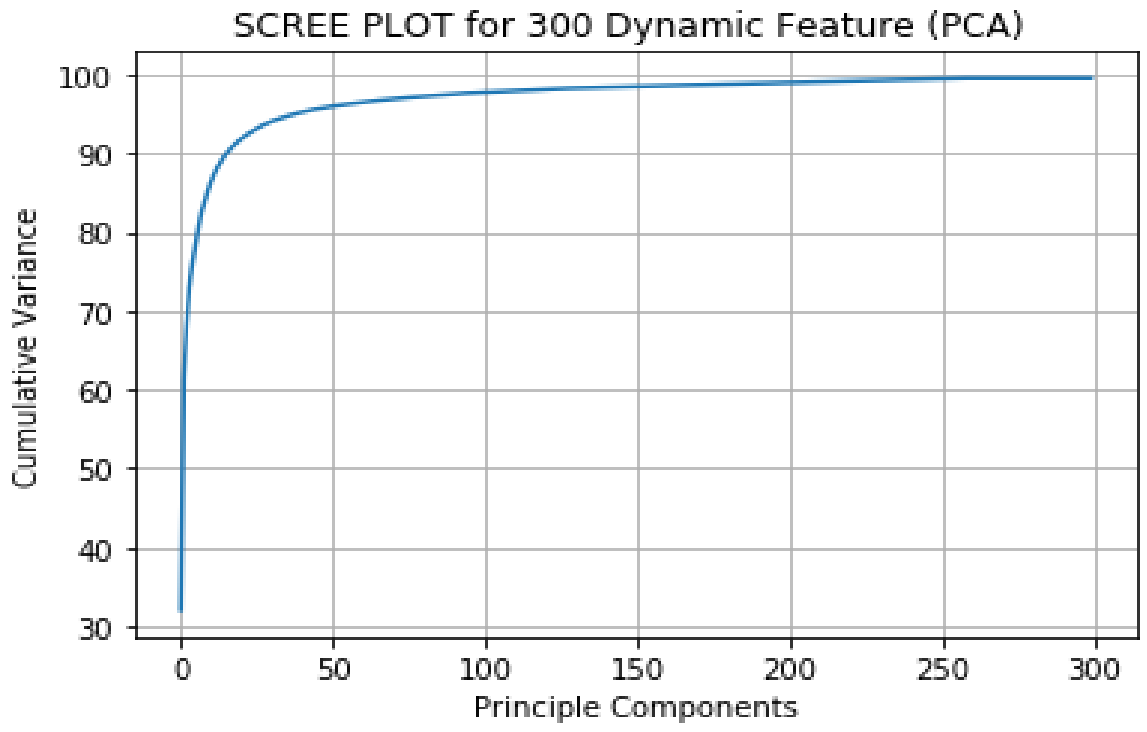

(a) **Figure 8.1 :Scree plot for static feature vector**

(b) **Figure 8.2: Scree plot for 300 dynamic feature**

**Figure 8 :Scree plot of static and dynamic data**

### 6.1.2 MI analysis of dynamic data

MI was applied on our dynamic dataset and scores were calculated for all dynamic features. Figure 11 shows, number of features vs achieved accuracy with the increase in features on x-axis and y-axis respectively. Accuracy was calculated using SVM, which shows maximum accuracy was achieved only in top 300 selected features.

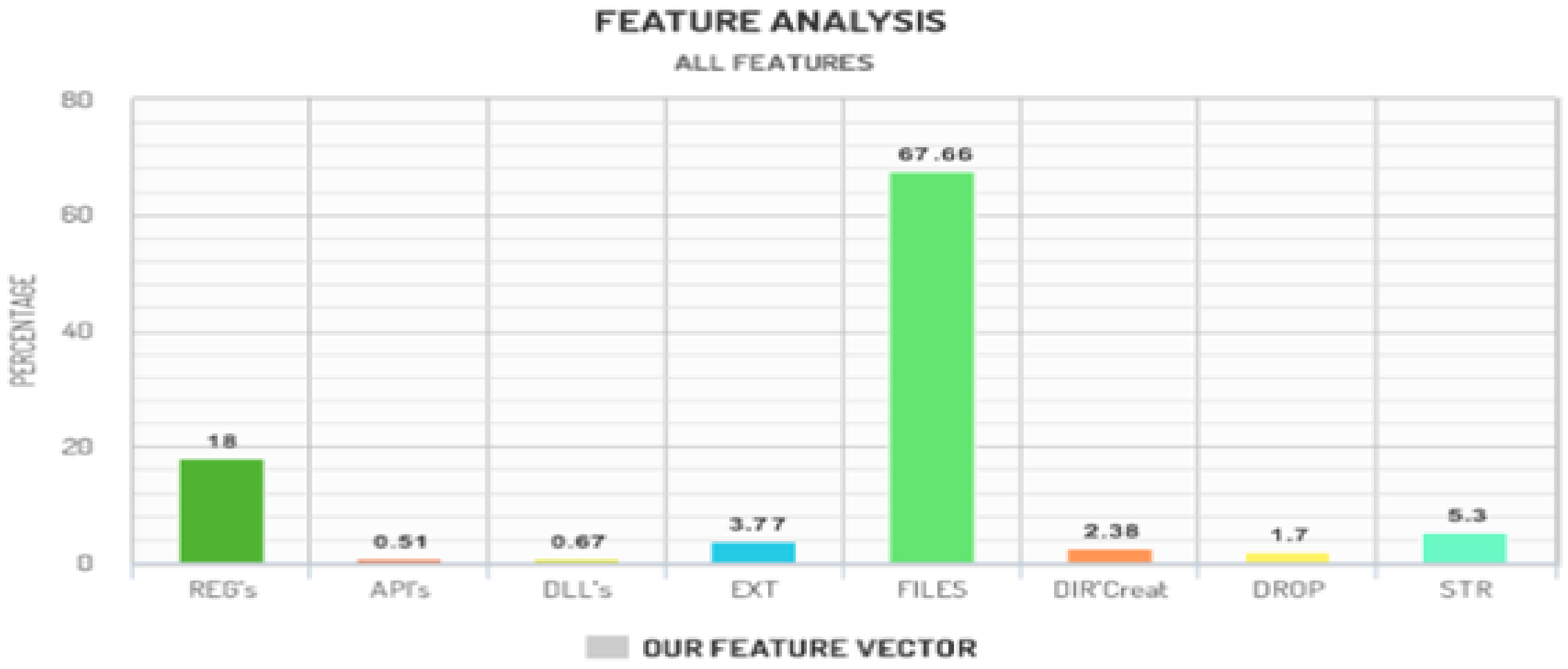

**Figure 9 :Feature vector visualization**

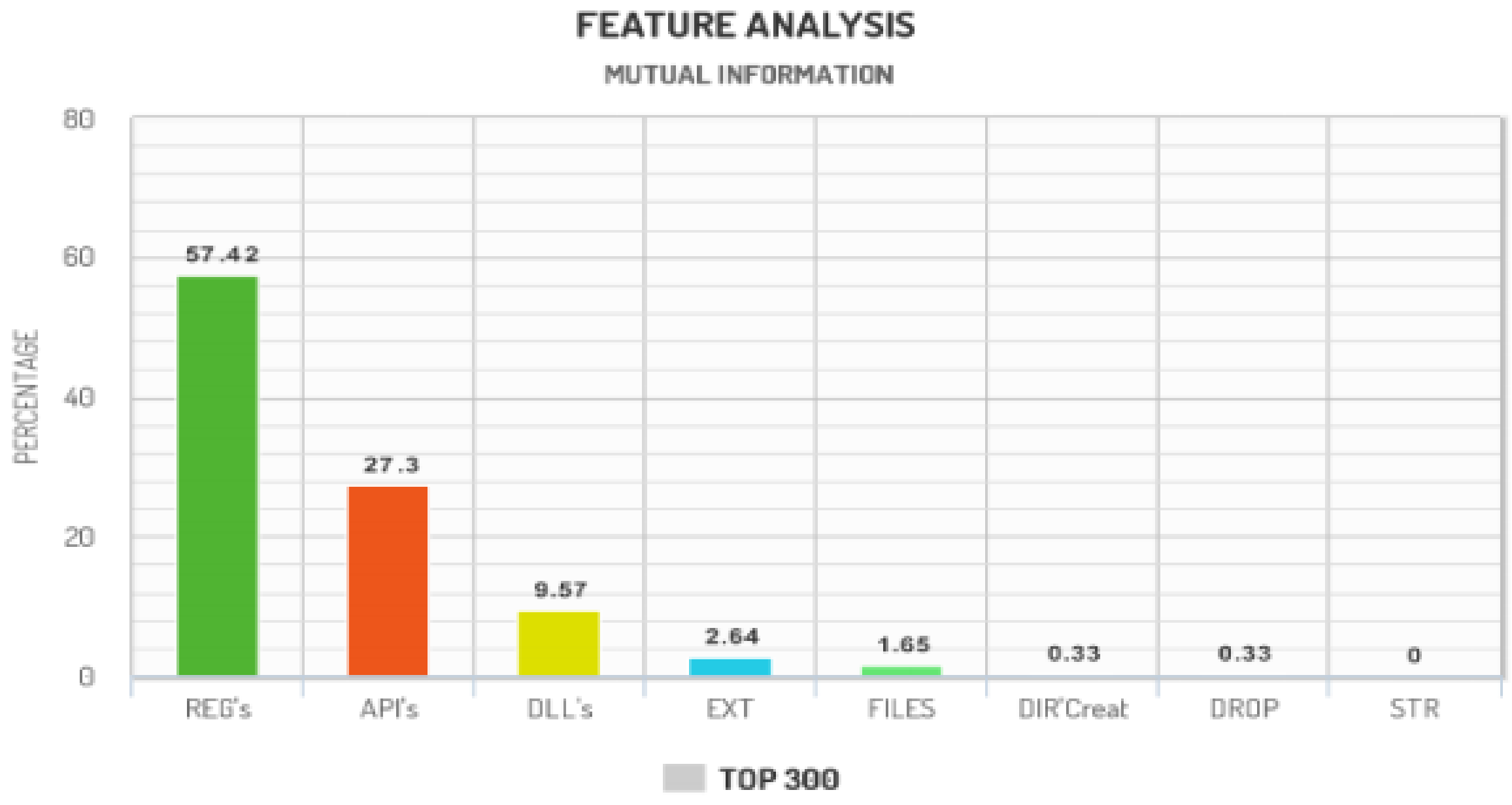

**Figure 10: Top 300 feature vector visualization**

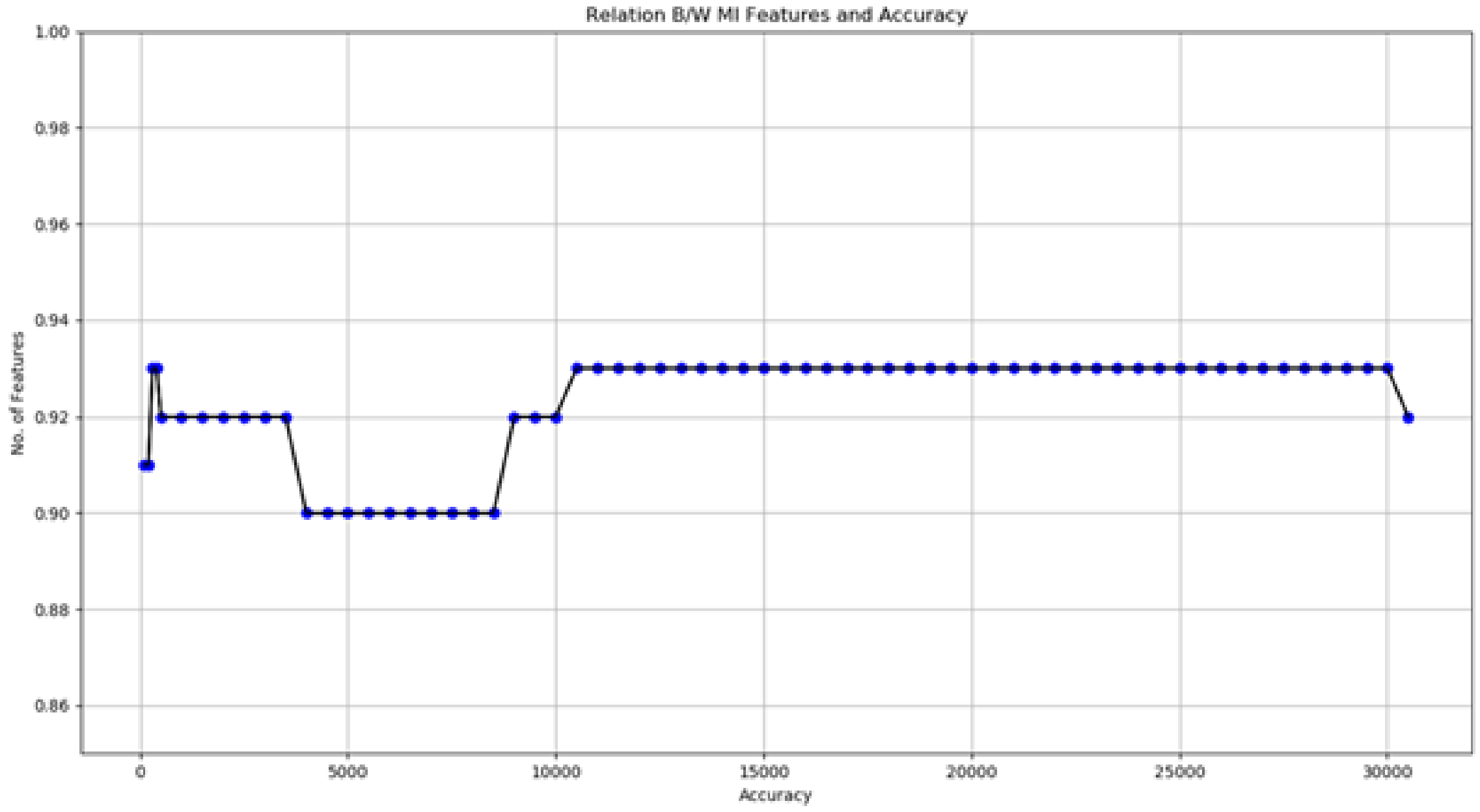

**Figure 11: Selected features vs accuracy**

Figure 9 shows the percentage of all recorded behavior termed as feature vector having length 47084, where one can see that number of files operation (READ, WRITE, OPEN, DELETE) has greater contribution, almost 67% of feature vector length(47084) than Registry Operations (READ, WRITE, OPEN, DELETE) which is only 18%. After applying MI Feature Selection technique, it is observed that only 300 features are enough to attain good accuracy as shown in Figure 10. Bar chart of selected top 300 feature vectors are shown in Figure 10. It can be observed that Files Operation which have greater contribution in feature vector of length 47084 are showing less contribution in selected top 300 feature vector that is nearly 1.65% respectively of total. Therefore, their importance is relatively low. However, Registry operations and API calls that were contributing nearly about 18% and 0.51% respectively, in feature vector of length 47084, but in selected top 300 feature vector they have played an important role by contributing 57% and 27 % respectively. Therefore, these Registry operations and API calls are very important to detect Ransomware behavior. After applying MI feature selection technique, the summary of above mentioned statement tells us that the selected top 300 feature vector has improved many folds.

#### 6.1.3 N-Gram technique analysis of dynamic data

MI results motivated us to determine the underlying hidden pattern that are handy in detection. For this purpose we further investigates the inside pattern and get valuable information for malware analyst. By the use of n-gram technique on registry deleted and registry created, it is observed that there are present some sequences of operations, which are always followed by malicious files but there is no order or sequence of these operations present in benign files. On Registry Delete operation if malicious file tries to delete registries, it follows a specific sequence. In our case N= (3 & 4) and this sequence repeated many times for every malicious files but for benign file it doesn't follow any specific sequence i.e., no repetitions in sequences. There was no common Registry deleted sequence between malicious and benign files.

### 6.2 Ransomware detection performance

This section, presents the test results by different trained classifiers over unseen static and dynamic datasets. It is important to notice that malicious files are named as positive files. Therefore, it is more important to notice that malicious files should not be predicted as benign. Therefore, false negative rate should be less. Satisfactory results obtained in testing are shown and discussed here. Detection performance on static and dynamic features is presented in Table 4 and Table 5 respectively. Figure 12 and Figure 13 are showing the area under the curve on validation and test set respectively.

In static dataset, based on best estimator in cross validation of Random Forest i.e. estimator = 800 testing the unseen static data got following results shown in Table 4: Achieved accuracy is 98% with only 0.03 false negative rate. Precision and recall are 99% and 96% respectively. Whereas, SVM is showing 96% accuracy 98% recall and 0.99 AUC (as shown in Figure 13). Although SVM has achieved less accuracy than Random Forest, but have more recall than Random Forest.

**Table 4: Static results**

| Model Names | Accuracy | Precision | Recall | F1-Score |
|---|---|---|---|---|
| **SVM** | 96% | 95% | 98% | 96% |
| **Random Forest** | 98% | 99% | 96% | 97% |

In dynamic dataset, best parameters selected through cross validation with SVM are C= 10, G = 0.013. On performing test over unseen data, accuracy gain by SVM is 92% (as shown in Table 5) with 0.009 false negative rate and AUC=0.92(as shown in Figure 13) . It can be observe that Random Forest is performing better than SVM in term of precision and F1 score, but SVM is good in predicting the positive class that is evident from the values of observed recall values that is 99%.

Overall from results it is evident that selected static features are performing well than the selected dynamic features. Moreover, Random forest is good in predicting the benign class and SVM is showing good performance in predicting positive class.

**Table 5: Dynamic results**

| Model Names | Accuracy | Precision | Recall | F1-Score |
|---|---|---|---|---|
| **SVM-RBF Kernel** | 92% | 85% | 99% | 91% |
| **Random Forest** | 91% | 98% | 86% | 92% |

For transfer learning weights of pre-trained ResNet-18 architecture on ImageNet were used. Criterion loss used in this model is Cross entropy loss. Optimizer used is Stochastic Gradient Descent. There are two main approaches to implement transfer learning weight initialization and feature extraction. We applied the Feature Extraction, transfer learning approach. Results given in Table 6 are showing that model is generalized well for unseen data and source domain has successfully transferred its knowledge to target domain.

**Table 6: ResNet-18 results on dynamic data**

| Model Names | Training Accuracy | Training Loss | Validation Accuracy | Validation Loss |
|---|---|---|---|---|
| **ResNet-18** | 94% | 0.17 | 92% | 0.23 |

**Static ROC Curve**

**Random Forest**

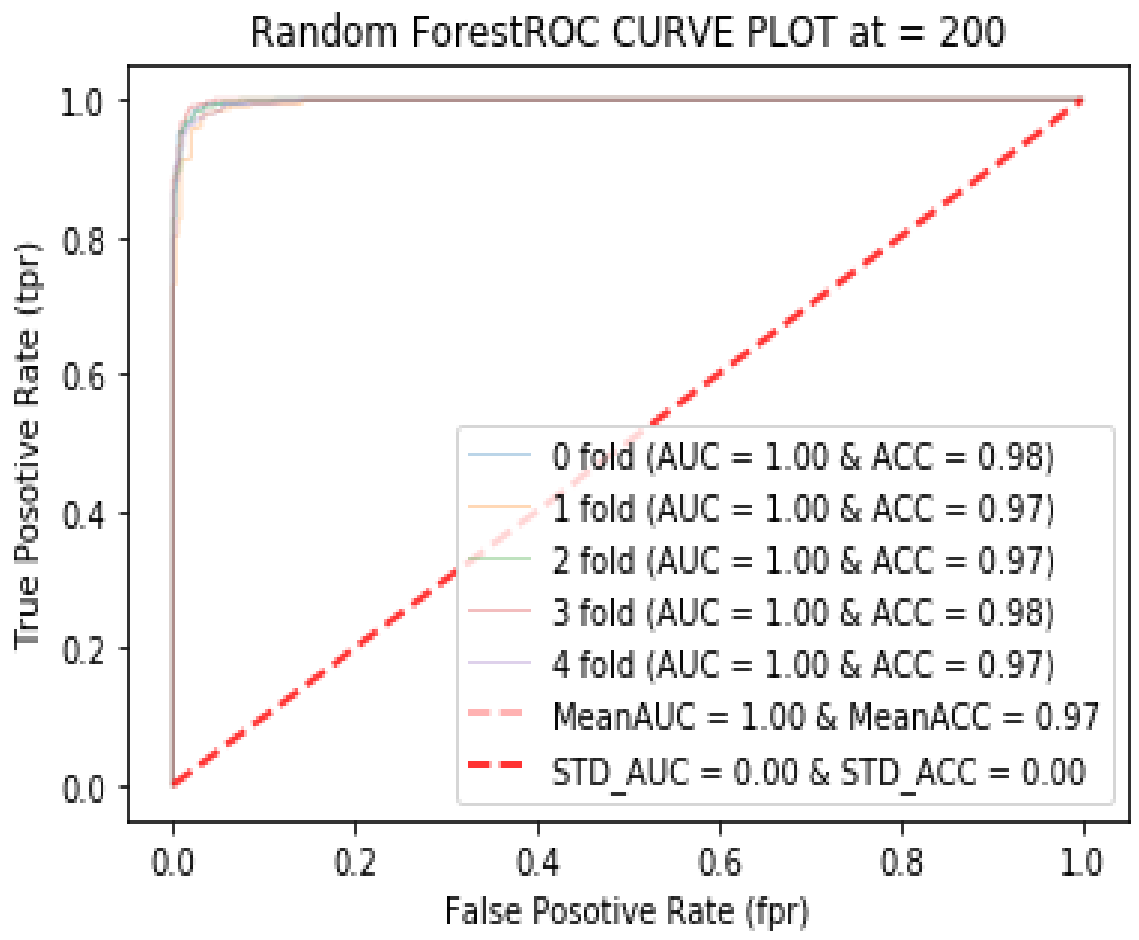

**Dynamic ROC Curve**

**Random Forest**

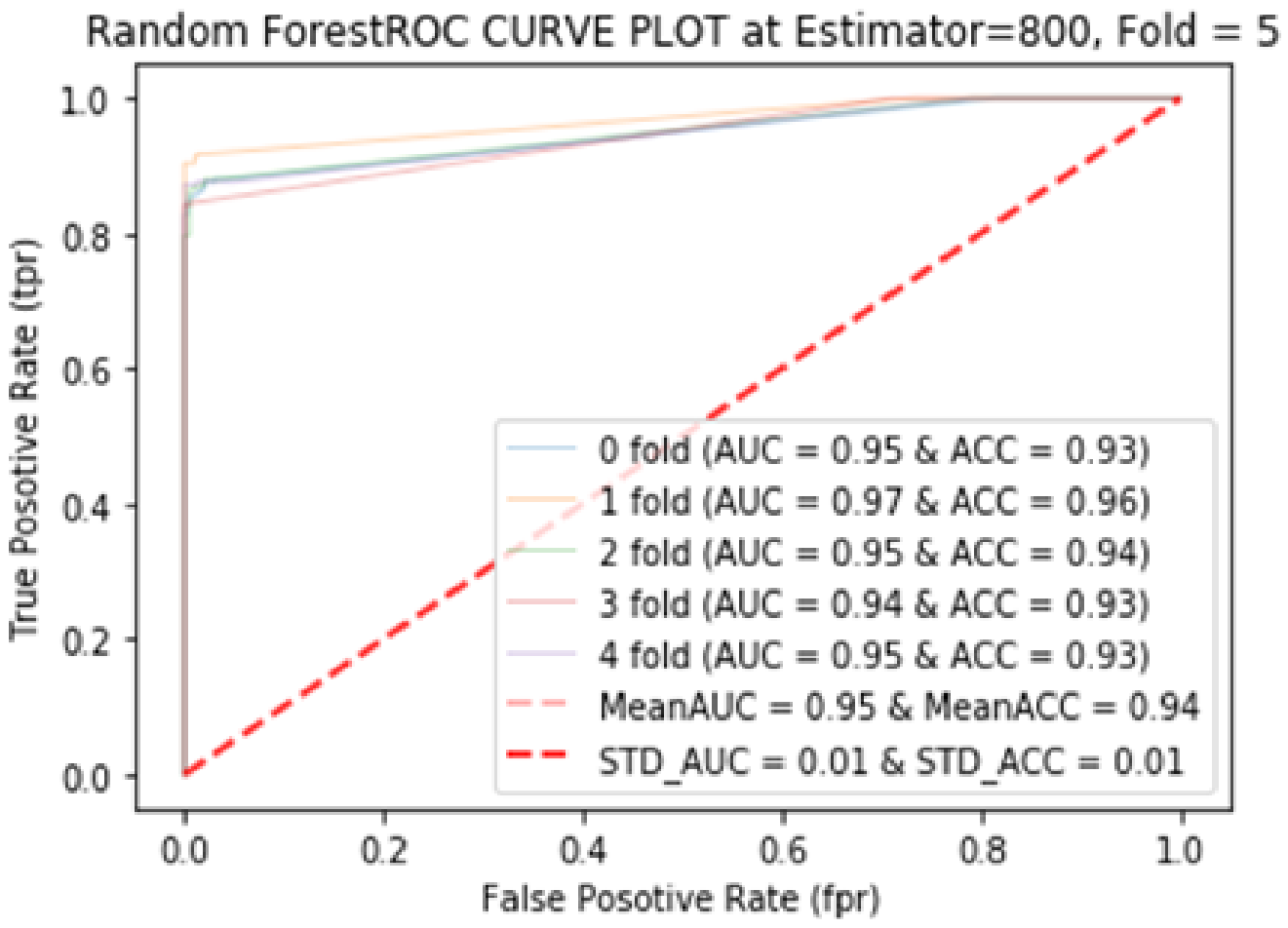

**SVM**

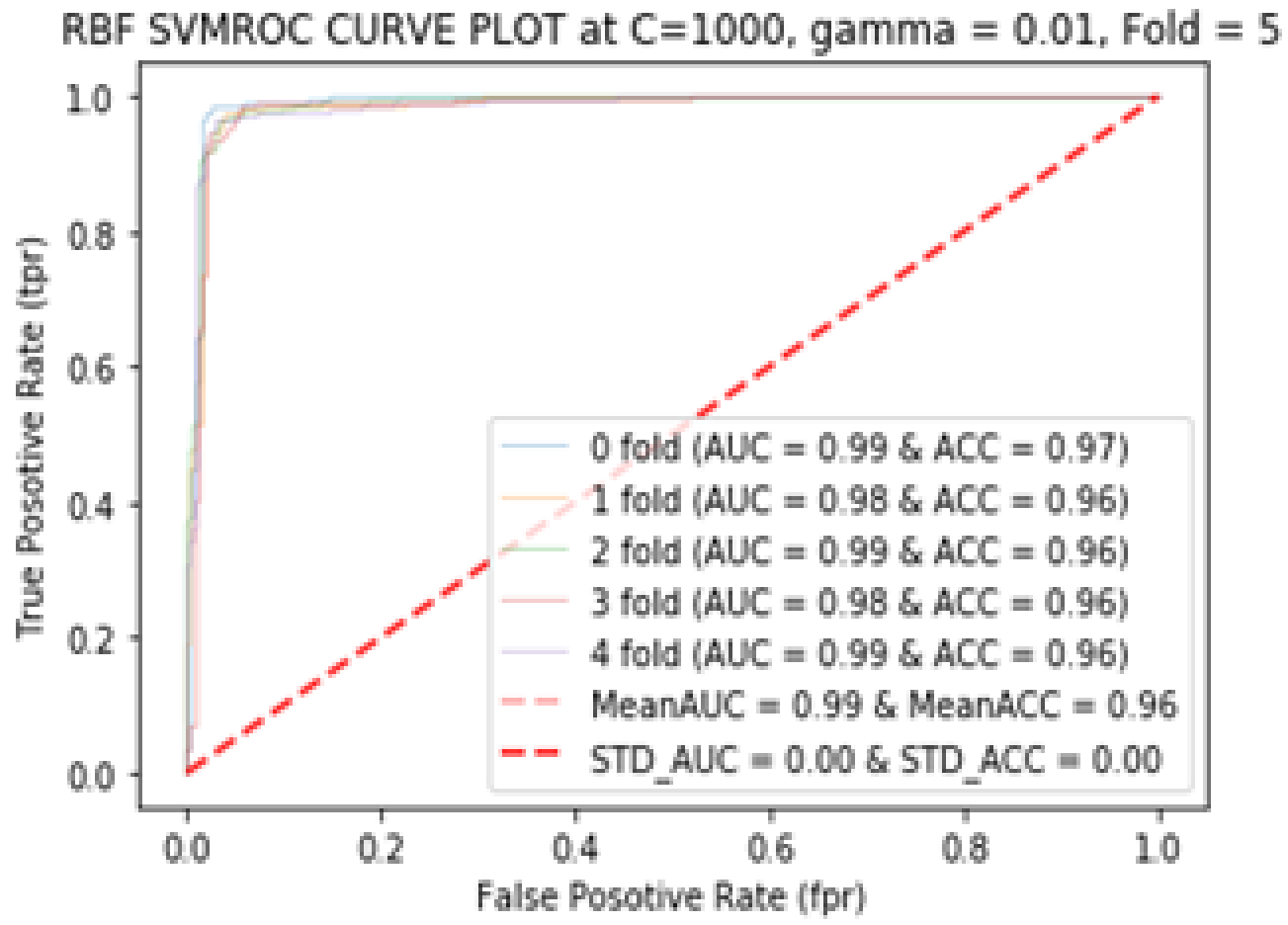

**SVM**

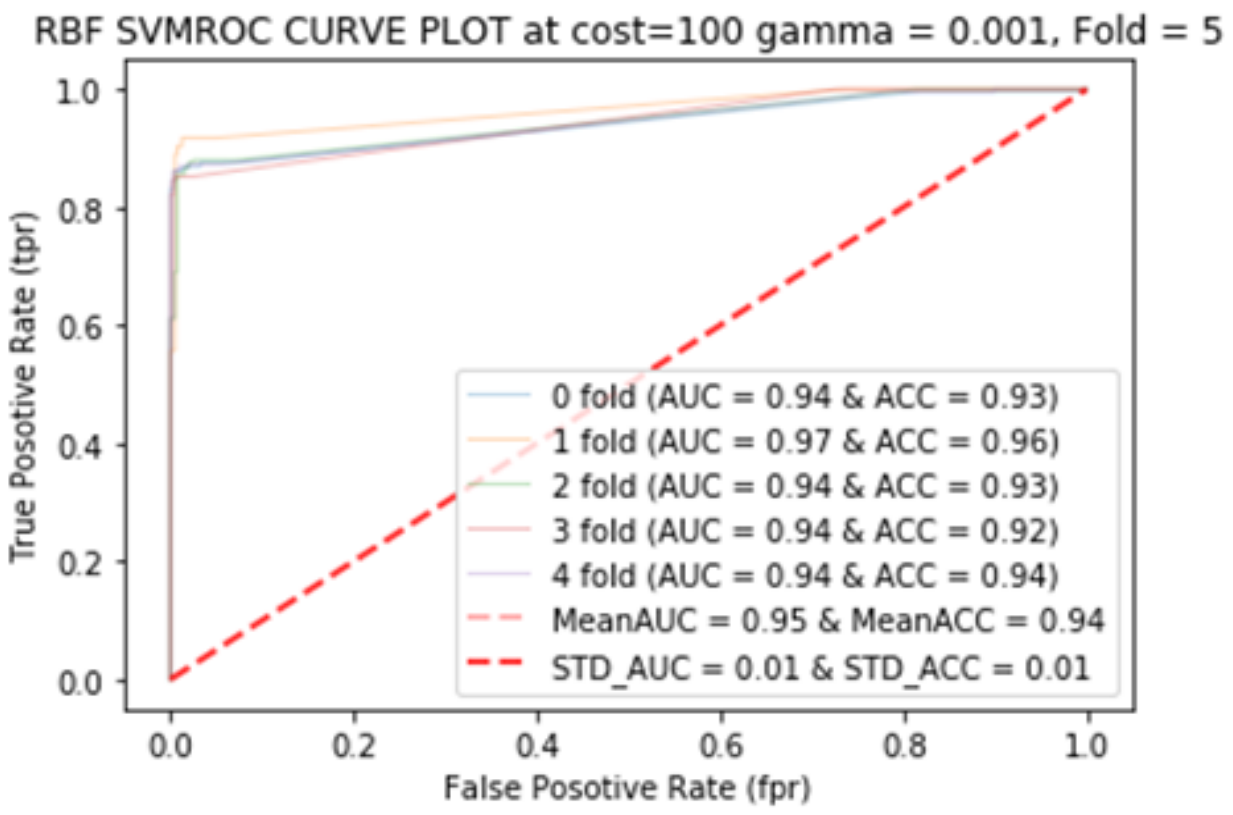

**Figure 12 :Results on validation set**

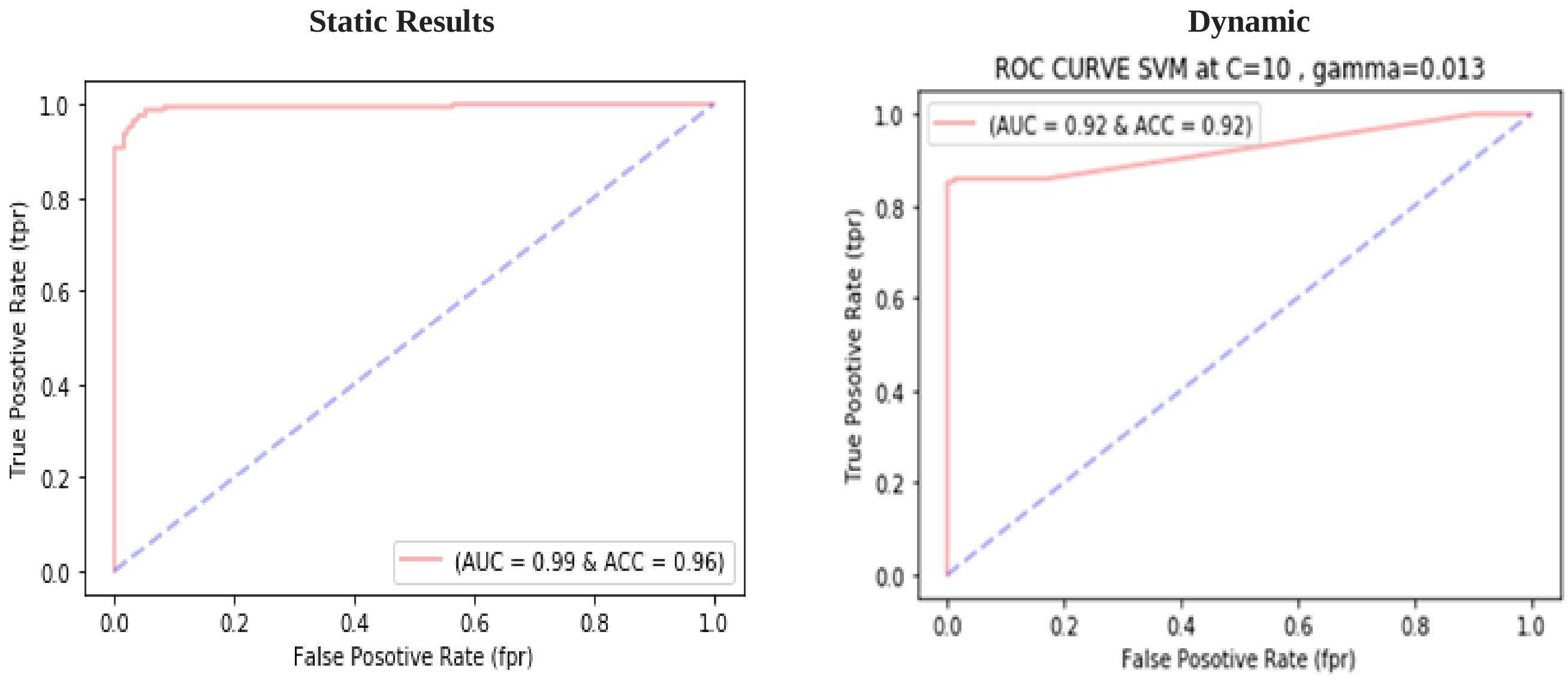

**Figure 13: Results of SVM on test set**

# 7 Conclusion

In the proposed RanSD technique, PE File and Cuckoo Sandbox are used to generate static and dynamic data respectively. In static analysis, physical properties of a file are used for analysis. PCA was applied over 89 features of static dataset which shows about 99% of variance was being captured in 56 dimensions. In Random-forest, accuracy achieved by static analysis is 98% with false negative rate of 0.03. Results over static analysis are very satisfactory but with the increasing attention of intruders and the way they change in code, solely static analysis is not sufficient and trust worthy. Here, dynamic analysis plays an important role in the detection of malicious files. Initially, by applying PCA 99% of variance was being captured in 300 dimensions of about 40,000+ dynamic features. To analyze the most important features MI was applied. That validates the results of PCA that 300 dimensions were useful in attaining 99% of variance. It reveals that maximum accuracy was achieved with 300 features but after that, accuracy dropped down due to presence of redundant features. Accuracy regains with the increase in number of features. Therefore, to avoid unnecessary computation it's preferable to take the top 300 features. These 300 features lead us to 92% of validation accuracy and 0.23 loss by the use of pre-trained ResNet-18 model. Sequences were found by the use of N-Gram method, which help us to examine the difference between the behavior of malicious and benign files. These sequences can be used in future with the help of Genetic Algorithms (Khan et al. 2019a) to improve the prediction performance and reduce computational time.